\documentclass[sigconf]{acmart}
\usepackage[utf8]{inputenc}
\usepackage{caption}
\usepackage{natbib}
\usepackage{color}
\usepackage{booktabs}
\usepackage{multirow}
\usepackage{pifont}
\usepackage{afterpage}
\usepackage{rotating}
\usepackage{amsmath}
\usepackage{textcomp} 
\usepackage{color}
\usepackage{array}
\usepackage{xcolor}
\usepackage{caption}
\usepackage{subcaption}
\usepackage{amssymb}
\usepackage{commath}
\usepackage{array}
\usepackage{flushend}
\newcolumntype{L}[1]{>{\raggedright\let\newline\\\arraybackslash}p{#1}}  

\usepackage{geometry}
\newcolumntype{x}[1]{>{\raggedright\arraybackslash}p{#1}}
\usepackage{makecell}
\usepackage{graphicx}
\usepackage{verbatim}
\usepackage{tikz}
\usepackage{mathtools}
\usepackage{arydshln}
\usepackage{hhline}
\usepackage{ragged2e}
\usepackage{enumitem}
\usepackage{url} 
\usepackage{mdframed}
\usepackage{xcolor}
\usepackage{tabularx}
\usepackage{CJKutf8}
\usepackage[T1]{fontenc}
\usepackage{balance}

\newcommand{\etc}{et al. }

\newcommand{\citetable}[1]{{Table}~\ref{#1}}

\copyrightyear{2020}
\acmYear{2020}
\setcopyright{acmcopyright}
\acmConference[MSR '20]{17th International Conference on Mining Software Repositories}{October 5--6, 2020}{Seoul, Republic of Korea}
\acmBooktitle{17th International Conference on Mining Software Repositories (MSR '20), October 5--6, 2020, Seoul, Republic of Korea}
\acmPrice{15.00}
\acmDOI{10.1145/3379597.3387440}
\acmISBN{978-1-4503-7517-7/20/05}

\begin{document}
\title{Traceability Support for Multi-Lingual Software Projects}
\author{Yalin Liu, Jinfeng Lin, Jane Cleland-Huang }
\affiliation{%
   \institution{University of Notre Dame}
   \city{Notre Dame}
   \state{IN}
}
\email{yliu26@nd.edu, jlin6@nd.edu, JaneHuang@nd.edu}


\begin{abstract}
Software traceability establishes associations between diverse software artifacts such as requirements, design, code, and test cases. Due to the non-trivial costs of manually creating and maintaining links, many researchers have proposed automated approaches based on information retrieval techniques. However, many globally distributed software projects produce software artifacts written in two or more languages. The use of intermingled languages reduces the efficacy of automated tracing solutions. In this paper, we first analyze and discuss patterns of intermingled language use across multiple projects, and then evaluate several different tracing algorithms including the Vector Space Model (VSM), Latent Semantic Indexing (LSI), Latent Dirichlet Allocation (LDA), and various models that combine mono- and cross-lingual word embeddings with the Generative Vector Space Model (GVSM). Based on an analysis of 14 Chinese-English projects, our results show that best performance is achieved using mono-lingual word embeddings integrated into GVSM with machine translation as a preprocessing step. 
\end{abstract}


\maketitle

\section{Introduction}
\label{introduction}
Software traceability establishes links between related artifacts such as requirements, design, code, and test cases \cite{DBLP:conf/icse/Cleland-HuangGHMZ14, DBLP:conf/re/GotelF94}. It provides useful supports for software engineering activities such as change impact analysis, safety assurance, test selection, and compliance analysis \cite{DBLP:journals/jss/MaderG12, RERoadmap}, and is prescribed in many safety-critical domains \cite{DBLP:conf/se/RempelMKC15}. The process of manually creating trace links is arduous and error-prone \cite{DBLP:journals/tse/HayesDS06} and therefore researchers have focused significant effort on automating the process through adopting diverse information retrieval techniques \cite{DBLP:journals/tosem/LuciaFOT07, DBLP:conf/icse/0002RGCM18, DBLP:journals/tse/HayesDS06, DBLP:conf/icse/0004CC17}. 
Automated techniques are of particular importance for reconstructing trace links after-the-fact, for example to fill in the gaps of an existing trace matrix, trace back to a new set of regulatory codes, or to establish trace links across existing Open Source Systems (OSS) for integration into safety-critical solutions \cite{SCS-OSS}. 

The most prevalent automated tracing techniques include the Vector Space Model (VSM) \cite{DBLP:journals/tse/HayesDS06}, Latent Dirichlet Allocation (LDA) \cite{DBLP:books/daglib/p/AsuncionT12}, Latent Semantic Indexing (LSI) \cite{DBLP:journals/tse/AntoniolCCLM02}, and deep-learning techniques \cite{DBLP:conf/icse/0004CC17}. Automated approaches typically analyze the textual content of each artifact, compute their syntactic and semantic similarity, and assign a relatedness score between each pair of artifacts that depicts the likelihood that the artifacts are associated. Despite the fact that industrial projects in involving international corporations often include artifacts with intermingled languages, research efforts have not addressed the challenge of intermingled languages.

We observed the problem of intermingled language use in a recent collaboration with an international corporation. Our goal was to help the company implement state-of-the-art automated tracing techniques across a repository of diverse artifacts; however, we found that many documents included a combination of two different languages (in this case English and Chinese). We then searched for other examples of intermingled language use in issues, requirements, and source code in OSS projects developed in common outsourcing countries, especially those that exhibited lower than average scores on the English Proficiency Index (EPI)\cite{epi}. 

This observation raised new questions about the effectiveness of different tracing algorithms when applied to artifacts containing two or more different languages. We hypothesized that existing trace algorithms that are effective in mono-lingual environments are not necessarily effective when applied to bilingual ones. We focused our efforts primarily on intermingled bilingual artifacts (IBAs) that included English and Chinese because our collaborators' project included these two languages, and because we found this to be the most prevalent language combination in OSS.

This paper, therefore, investigates the automatic generation of trace links within software projects with IBAs, which we formally specify as follows. Given a dataset of artifacts $D$ with source artifact set $A_S$ and target artifact set $A_T$, then source artifact $a_{s_i} \in A_S$ is composed of terms in a vocabulary V, where $V = L_p \cup L_f $. The target artifacts are constituted similarly to the source artifacts. Further, $L_p$ and $L_f$ are vocabulary subsets of primary language and foreign language. 

The remainder of the paper is laid out as follows. Section \ref{sec:datasets} describes the datasets that were used throughout the remainder of the paper. Section \ref{sec:IBAPractice} analyzes the usage of intermingled language across 10 different Chinese-English projects and identifies commonly occurring usage patterns.  Section \ref{sec:trace_algorithms} describes three classic tracing algorithms -- namely the Vector Space Model (VSM), Latent Dirichlet Allocation (LDA) and Latent Semantic Indexing (LSI), and evaluates their accuracy in multilingual project environments with and without the use of translation as a preprocessing step.  Section \ref{sec:approach} then introduces the Generalized Vector Space Model (GVSM) and evaluates its effectiveness in combination with both mono- and cross-lingual word embeddings. We show that utilizing mono-lingual word embeddings with a preprocessing translation step tends to be more effective than the use cross-lingual embeddings; however, to avoid the costs of an external translation service, an individual corporation might opt for the cross-lingual approach which is easier to train than a language translator. Finally, in Sections \ref{sec:threats} to \ref{conclusion}, we discuss applications to other languages, threats to validity, related work, and finally conclude by summarizing our results and discussing their relevance.

\section{Experimental Datasets}
\label{sec:datasets}
To establish the experimental environment used throughout the experiments described in this paper, we collected a dataset of 17 OSS projects, each containing artifacts written in English plus one additional language. We refer to this second language as the \emph{foreign} language. All projects met the following criteria:
\begin{enumerate}
    \item The Project contains at least 40 issues and commits in its overall development history.
    \item Foreign terms constituted at least 1\% of the vocabulary.
    \item Tags were routinely created to include issue IDs in the commit messages (i.e., to generate trace links for evaluation purposes).
    \item The project exhibited diversity in size of links between issues and commits, in comparison to other selected projects. This enabled us to observe the performance of our model in both large and small projects.
\end{enumerate}

\begin{table}[t!]
	\centering
	\caption{ OSS projects from Github used in our study}
	 \addtolength{\tabcolsep}{-3.5pt}
        \begin{tabular}{|l|l|l|l|L{2.5cm}|}
        \hline
        {\bf Lang} &  {\bf Project } &  {\bf Abbrv.} &  {\bf Comp.}& {\bf Domain } \\ \hline
        \multirow{15}{*}
         { Chinese } & Arthas & Ar &Alibaba&Java diagnostics \\ 
        & bk-cmdb & BK &Tencent&Config. Manage. \\
        & Canal & Ca &Alibaba&Database log parser \\ 
        & Druid & Dr &Alibaba&Database connect. \\ 
        & Emmagee & Em &Netease&Performance test \\
        & Nacos & Na &Alibaba&Service discovery  \\ 
        & NCNN & Nc &Tencent& Neural network \\
        & Pegasus & Pe &Xiaomi& Storage system \\ 
        & QMUI\_Android & QMA &Tencent&Mobile UI \\
        & QMUI\_IOS & QMI &Tencent&Mobile UI \\
        & Rax & Ra &Alibaba& Application builder \\  
        & San & Sa &Baidu& JavaScript comp. \\ 
        & Weui & We &Tencent&Mobile UI \\
        & xLua & xL &Tencent&Programming \\ \hline
        Korean & Konlpy & Ko &Personal& NLP package \\ \hline
        Japanese & Cica & Ci &Personal& Font repository \\ \hline
        German & Aws-berline &Ab&Personal&  Website \\\hline
        \end{tabular}
	\vspace{12pt}
	\label{tab:project_summary}
\end{table}
\begin{table}[ht]
	\centering
	\vspace{-12pt}
	\caption{An example of IBA artifacts. In this case, the commit message, issue summary, and commit content all contain foreign terms intermingled with English ones.}
	 \small
    \begin{subtable}[h]{\columnwidth}
        \centering
        \begin{tabular}{|L{1.5cm}|p{6.3cm}|}
        \hline
        Commit ID & 2017fb7cf12c... \\\hline
        Commit message & \begin{CJK*}{UTF8}{gbsn}
        PagerUtils offset的bug，当 0 ti需要修改为0时，取值不正确 \end{CJK*}\\ \hline
        Change set & 
        [-] if (offset > 0) \{\newline
        [+] if (offset >= 0) \{\newline
        [+] \begin{CJK*}{UTF8}{gbsn}//测试mysql 4\end{CJK*} \newline
        [+] public void test\_mysql\_4() throws Exception \{\newline
            \ldots 
       \\ \hline
        
       \end{tabular}
       \caption{The commit message and its change set served as the \emph{source} artifact. + sign (-sign) refer to added (deleted) content in a commit}
       \label{tab:commit_example}
    \end{subtable}
           \begin{subtable}[h]{\columnwidth}
        \centering
        \begin{tabular}{|L{1.5cm}|p{6.3cm}|}
        \hline
        Issue ID & Issue \#3428 \\\hline
        Summary & \begin{CJK*}{UTF8}{gbsn}
        缺少java.sql.Time类型适配 \end{CJK*}\\ \hline
        Description& 
        - \begin{CJK*}{UTF8}{gbsn} '2019-08-29 13:54:29.999888'这个为啥是6位，不是3位么? \end{CJK*}\newline
        - MySQL 5.7 has fractional seconds support for TIME, DATETIME, and TIMESTAMP values, with up to microseconds (6 digits) precision. \begin{CJK*}{UTF8}{gbsn} 因为这里的sql仅仅是用来看看的，不能拿去数据库执行，如果要执行的话还得考虑mysql时区与程序时区的问题\newline
            \ldots 
            \end{CJK*}
       \\ \hline
       \end{tabular}
       \caption{The issue including its description and subsequent discussion served as the \emph{target} artifact.}
       \label{tab:issue_example}
    \end{subtable}
    \vspace{-14pt}
    \label{tab:commit-issue}
\end{table}

To identify datasets meeting these criteria we (1) collected the names of top Chinese IT companies based on a survey published by Meeker \etc \cite{meeker2018internet}. Nine Chinese companies, including Alibaba, Tencent, Meitaun, JD, Baidu, NetEase and XiaoMi, were recognized amongst the top 30 global IT companies and were included in our project search.
(2) We searched Github using these company names to retrieve a list of their open source repositories. We found enterprise-level open source repositories for eight of the nine companies (excluding JD). (3) We then sorted the projects in the retrieved repositories by the numbers of stars to identify the most active projects. (4) Finally, we selected the top scoring projects that met our defined criteria.  In addition, while our focus throughout this paper is on English-Chinese projects, we also included three additional projects based on Korean, Japanese, and German. However, as large companies in those three countries, e.g. Samsung, Hitachi and Siemens, produce few bilingual projects, we selected three popular personal projects in those languages instead. We searched by the language name, and then followed steps (3) and (4) as described above. We discuss results from these additional languages towards the end of the paper. 
The selected projects are depicted in Table \ref{tab:project_summary}.

\begin{table*}[ht]
	\centering
	\caption{OSS datasets used for experiments showing counts of artifacts and links, and percentages of foreign terms. 14 Chinese datasets are shown on the left, and three non-Chinese datasets on the right.}

	 \small\addtolength{\tabcolsep}{-1pt}

    \begin{subtable}{\linewidth}
            \begin{tabularx} {\linewidth}{*{15}c | *{3}c{>{\centering\arraybackslash}X}}
        	\hline
        	\textbf{Project Name} & Ar & Bk & Ca & Dr & Em & Na& NC & Pe & QMA & QMI & Ra & Sa& We & xL & Ko & Ci & Ab \\ \hline
                Issue &437&1701&1080&2859&106&303&746&254&483&478&846&46&752&520&241&49&107 \\
                Commit &489&4504&718&5840&139&471&568&261&296&464&3340&1426&507&741&503&188&299 \\
                Links & 167&1183&274&1173&32&161&101&163&71&35&573&276&159&53&33&27&75 \\
                Foreign Terms & 11.0\%&7.6\%&4.3\%&6.7\%&19.5&1.0\%&29.1\%&35.8\%&15.5\%\%&19.4\%&8.5\%&4.0\%&6.0\%&30.0\%&2.9\%&11.0\%&14.0\% \\ 
                \hline \hline
    	\end{tabularx}
    	\caption{Artifact counts for each dataset as mined from the OSS}    
    	\label{tab:data_status_1}
    \end{subtable}
	
	\begin{subtable}{\linewidth}
		\begin{tabularx} 
		{\linewidth}{*{15}c | *{3}c{>{\centering\arraybackslash}X}}
		\hline
        	  \textbf{Project Name} & Ar & Bk & Ca & Dr & Em & Na & NC & Pe & QMA & QMI &Ra & Sa & We & xL & Ko & Ci & Ab\\ \hline
                Issue & 122&895&232&1092&31&132&97&160&70&32&560&186&154&52&32&25&74 \\
                Commit & 167&1178&273&1161&32&161&99&160&71&35&571&275&159&52&33&27&74 \\
                Links & 167&1179&273&1161&32&161&99&160&71&35&571&275&159&52&33&27&74 \\
                Foreign Terms & 14.6\%&8.3\%&5.4\%&7.3\%&21.9\%&1.0\%&28.0\%&35.3\%&16.8\%&20.8\%&9.0\%&8.2\%&7.1\%&29.5\%&7.0\%&11.7\%&31.0\% \\ \hline \hline
    	\end{tabularx}
	\caption{Artifact counts following pruning to remove artifacts that are not impacted directly, or indirectly, by IBA.}
    	\label{tab:data_status_2}
	\end{subtable}
	\vspace{-12pt}
	\label{tab:data_status}
\end{table*}

We used the Github Rest API to parse each of the selected projects and to extract information from commits and issues. We retrieved both the \textbf{commit message} and the \textbf{source code change set} to establish source artifacts for the trace links.  We then collected \textbf{issue discussions}, and \textbf{issue summaries} to construct our target artifact sets. We removed personal identifications from all issues, while retaining comments.  
By selecting issues and commits for our tracing artifacts we were able to automatically establish a golden link set by using regular expressions to parse commit messages and extracting explicitly defined connections between commits and issues. An example of a commit and issue is depicted in Table \ref{tab:commit-issue}, and statistics for the collected datasets are shown in \citetable{tab:data_status_1}.

Rath et al., studied commit messages in five OSS projects and found that an average of 48\% were not linked to any issue ID \cite{DBLP:conf/icse/0002RGCM18}. This implies that our golden link set is likely to be incomplete and that `true positive' instances in the evaluation step could be mistakenly treated as `false positives'. To partially mitigate this problem, we limited the scope of artifacts to those included in at least one of the explicitly specified links from the golden link set. All other artifacts (i.e., issues and commits) were removed. This created a dataset with denser ground-truth link coverage and fewer inaccuracies. Furthermore, we decided to remove all artifacts that were not impacted, directly or indirectly, by the IBAs. For each link in the golden artifact set, if at least one artifact associated with that link included a foreign language term, then we retained both artifacts and labeled the link as an intermingled link. All other artifacts were removed. In effect, it removed artifacts that were never involved in any intermingled link, and allowed us to focus the evaluation on artifacts and links that were impacted by the use of foreign language terms. Applying these two pruning steps reduced the pruned dataset to an average of approximately 27\% of the original issues, 17\% of the commits, and 77\% of the links as shown in Table \ref{tab:data_status_2} \footnote{{Dataset can be found at http://doi.org/10.5281/zenodo.3713256}}.

For all tracing experiments using these datasets, we applied a time-based heuristic proposed by Rath et al.\cite{DBLP:conf/icse/0002RGCM18}. This heuristic states that as $Issue_{create} < Commit_{create}< Issue_{close}$, then commits can only be linked to currently open issues, as closed issues are unlikely to represent a valid trace link.

\begin{table}[h!]
	\centering
	\caption{Inductive open coding tags for 50 issues and 50 commits}
    \begin{subtable}{\columnwidth}
        \addtolength{\tabcolsep}{-3.4pt}
        \begin{tabular}{|l|L{1.7cm}|L{4.8cm}|l|l|}
        \hline
        {\bf Tag} & {\bf Usage}& {\bf Examples}&{\bf Eng}&{\bf Ch}  \\ \hline
        ID&Issue summary& Primary language of issue summary&24&76 \\ \hline
        IA&Issue description & Primary language of issue description &4&98\\ \hline 
        CM&Commit message&Primary language of commit message &86&14 \\\hline
        \end{tabular}
        \caption{Tags used to label the dominant language of the artifacts}
    \end{subtable}

        \begin{subtable}{\columnwidth}
            \addtolength{\tabcolsep}{-3.4pt}
            \begin{tabular}{|l|L{1.9cm}|L{4.5cm}|l|l|}
            \hline
            {\bf Tag} & {\bf Usage}& {\bf Examples}&{\bf Eng}&{\bf Ch}  \\ \hline
            CR&Ext. reference& External system e.g., Tomcat, dashboard&56&0 \\ \hline
            V&Verb usage&Verbs from non-dominant language e.g., kill, debug&9&0\\ \hline
            T&Noun usage&Common objects from non-dominant language e.g., demo, thread, timestamp, \begin{CJK*}{UTF8}{gbsn}资源池\end{CJK*} for resource pool&36&6 \\\hline
            ER&Errors and traces & Error messages and stack traces &10&7 \\\hline
            AC&Acronym& \begin{CJK*}{UTF8}{gbsn}报错 ~= 报告错误\end{CJK*}; PR = pull request&9&0\\ \hline
            TAG&Tag use & [feature request],\begin{CJK*}{UTF8}{gbsn} [中文说明](README\_CN.md) \end{CJK*} &6&9\\ \hline
            CD&Code snippets&\begin{CJK*}{UTF8}{gbsn}println("代码植入成功");\end{CJK*}&28&19 \\\hline
            CC&Code comments& Comments in natural languages &0&68\\\hline
            $BD$   & Bilingual Duplication & Duplicated content written in two languages& 2&2	\\ \hline
            \end{tabular}
            \caption{Tags used to label roles of non-dominate phrases in a dominate language sentence}
    \end{subtable}
    \vspace{-18pt}
    \label{tab:tags}
\end{table}

\section{Multilingual Artifacts in Practice}
\label{sec:IBAPractice}
To lay a foundation for our work, we first investigate how terms from two different languages are intermingled across issues and commits in projects where both English and Chinese are used. Our first research question is therefore defined as follows: 
\begin{itemize}[leftmargin=*]
\item {\bf RQ1:} How are Chinese and English terms intermingled across different artifacts?
\end{itemize}

\subsection{Approach} First, we applied stratified random selection to collect 5 issues and 5 commits from each of the first 10 projects listed in Table.~\ref{tab:project_summary}, producing an overall dataset of 50 issues and 50 commits. To analyze the Chinese and English usage patterns we adopted an inductive open coding technique\cite{khandkar2009open}.  
For each artifact, we first determined whether it was primarily written in Chinese or English (i.e., its dominant language). We then identified all phrases not written in the dominant language, and manually evaluated the role of those phrases within the artifact. Based on these observations, we created distinctive tags to categorize the discovered roles and used these tags to mark up each issue and commit artifact. We report results in Table \ref{tab:tags}.

\subsection{Observations}
Observing the combinations and occurrence of these tags, enables us to infer how Chinese and English languages were intermingled in the analyzed artifacts. In the following discussion we use the tags reported in \citetable{tab:tags} annotated with subscripts for Chinese (C) and English (E). 

\begin{itemize}[leftmargin=1.5em]
    \item 87\% of our analyzed issues were tagged with ID$_C$ or IA$_C$, meaning that in our datasets, the majority of issues were primarily described and answered in Chinese.
    \item While 28\% of issues were tagged with ID$_E$ or IA$_E$, we did not observe any cases in which they also included Chinese words.
    \item In contrast, ID$_C$ and IA$_C$ were combined with most types of English tags listed in Table.~\ref{tab:tags}. Chinese sentences were frequently intermingled with English terms.
    \item 86\% of the commits are tagged as CM$_E$, meaning that commit message were likely to be written in English and 14\% in Chinese.
    \item 68\% of commits were tagged with CC$_C$ and 38\% with CD$_C$ because Chinese frequently appeared in code comments and source code especially where database query and UI elements were discussed such as SQL query condition, output messages and UI widget labels.
    \item For the commits contain Chinese Commits (with CC$_C$ tag), 58\%, 51\%, 6\% and 6\% of them are also tagged with CR$_E$, T$_E$, V$_E$ and TAG$_E$. It indicates that Chinese comments are also likely to intermingled with English phrase, in which CR and T are the most common intermingle scenarios.
\end{itemize}

To summarize, we observed that Chinese sentences tended to include intermingled English phrases in both issue and source code, while English sentences rarely included Chinese phrases. Finally, the predominant role of a secondary language within the context of the primary language was to reference components or to use specific terminology. 
In related work, Timo \etc \cite{pawelka2015code}, investigated 15 OSS projects in which English was intermingled with other European languages. They found both identifiers and comments written in a second language were intermingled with source code. This differs from our observations of Chinese-English projects, in which  we only found comments, but not identifiers, written in Chinese or pinyin (an alphabetic representation of Chinese characters).

\section{Basic Translation Approaches}
\label{sec:trace_algorithms}
%
For our first series of traceability experiments, we utilize three commonly adopted tracing algorithms with and without the use of a basic translation step. This preliminary research question is an important one, because it addresses the question of whether the IBA issue can be addressed simply through applying a preprocessing translation step. We explore the improvement obtained by leveraging neural machine translation (NMT) as part of the tracing process to address the following research question:
\begin{itemize}[leftmargin=*]
\item {\bf RQ2:} To what extent does the use of a neural machine translation (NMT) approach improve the traceability performance of VSM, LDA and LSI for an IBA dataset?
\end{itemize}

\subsection{Baseline Algorithms}
The Vector Space Model (VSM), Latent Dirichlet Allocation (LDA) and Latent semantic indexing (LSI) are three models commonly used to generate trace links. Researchers have successfully applied those models for various tracing scenarios within mono-lingual environments \cite{6341764,lormans2006can,asuncion2010software}. However, it is unknown whether those models are effective for IBA datasets. In this section, we therefore describe these three common trace models.  

\subsubsection{Vector Space Model}
\label{sec:trace_algorithm_vsm}
VSM is one of the simplest techniques for computing similarity between documents and has been used to support diverse trace link recovery tasks \cite{DBLP:journals/tosem/LuciaFOT07,DBLP:journals/tse/HayesDS06}. Despite its simplicity it has been shown to frequently outperform other techniques across many different datasets \cite{DBLP:conf/sigsoft/LoharAZC13}. VSM represents the vocabulary of discourse as an indexed linear vector of terms, while individual documents (aka artifacts) are represented as weighted vectors in this space. Weights are commonly assigned using TF-IDF (Term frequency-inverse document frequency) in which the importance of a term is based on its occurrence and distribution across the text corpus. VSM assumes the Bag of Words (BOW) model in which the ordering of words is not preserved.  Let $A_S$ be the collection of source artifacts and $A_T$ the collection of target artifacts then each artifact $a_i \in A_S\cup A_T$ is represented by the terms $\{t_1...t_n\}$ it contains regardless of their order. Each artifact $a_i$ is transformed into a numeric format $a_i = \{w_1,w_2,\dots,w_n\}$ where $w_n$ indicates the TF-IDF score for $t_i$. The similarity of two artifacts is then estimated by measuring the distance between their vector representations -- often by computing the cosine similarity between source and target vectors as follows:
\vspace{-4pt}
\begin{equation}\label{VSM_EQ}
        Similarity(a_i,a_j) = \frac{a_i^T  \cdot a_j}{\sqrt{a_i^T \cdot a_i}\sqrt{a_j^T \cdot a_j}}
\end{equation}

\vspace{-4pt}
From the perspective of VSM, words are indexed as atomic units and are orthogonal to each other regardless of their semantic similarity. Therefore, the affinity between two artifacts is evaluated based on the volume and quality of their common vocabulary. IBA datasets, in contrast to mono-lingual ones, have a richer vocabulary composed from two different languages; therefore, source and target artifacts could be semantically similar, yet written using terminology from two distinct languages. This consequently could lead to an underestimation of the artifacts' affinity.

\subsubsection{Topic Modeling Approaches}
Topic modeling is also frequently used to support automated trace link generation \cite{DBLP:conf/icse/AsuncionAT10}.Topic modeling techniques discover the hidden semantic structures of artifacts as abstract concepts and then represent the artifact as a distribution over the concepts. The most commonly adopted approaches are Latent Dirichlet Allocation (LDA), Latent Semantic Indexing (LSI) and Probabilistic Latent Semantic Indexing (PLSI). LSI, also known as Latent Semantic Analysis (LSA), represents each artifact $a_i$ as a vector of word counts $c_n$ such that each word is represented as $a_i= \{c_1,c_2,...,c_n\}$. Thus the artifact corpus $A$ can be represented as a matrix $A =\{a_1,a_2,...,a_m\}$ where $m$ refers to the total number of all artifacts in $A$. LSI learns the latent topics by applying matrix decomposition, e.g Singular Value Decomposition (SVD) \cite{DBLP:journals/tosem/LuciaFOT07}. Hofmann \etc proposed a probabilistic variant of LSI which is known as PLSI in 1999 \cite{DBLP:conf/sigir/Hofmann99} in which a probabilistic model with latent topics is leveraged as a replacement of SVD. LDA then can be regarded as a Bayesian version of PLSI where dirichlet priors are introduced for the topic distribution. Given the topic distribution vector of source and target artifacts, the affinity between two artifacts can be calculated either with Cosine similarity or with Hellinger distance \cite{1089532} which quantifies the similarity between two probability distributions.
As we know, topic modeling methods represent each topic by eliciting a group of distinctive words associated with a similar theme. The per-artifact topic distribution probability, indicating the affinity between the artifact and a topic, is obtained by analyzing the artifact vocabulary and those selected words. When project artifacts contain foreign languages, the representative topic words are constituted from two (or more) distinct languages. Topic modeling methods therefore face similar vocabulary challenges to VSM. The use of foreign languages introduces a new set of words from a different language and thereby reduces the likelihood that related artifacts use the same words.

\subsection{Leveraging Neural Machine Translation}
\label{sec:impact_of_binlingual}

Neural machine translation services, such as Google Translate, are capable of translating documents with complex grammars into diverse languages. Wu \etc demonstrated that, for a specific set of datasets, Google Translate achieved the average accuracy of bilingual human translators. Furthermore, current versions of Google Translate have addressed approximately 60\% of known translation errors in popular languages such as English-Chinese, thereby significantly improving performance \cite{DBLP:journals/corr/WuSCLNMKCGMKSJL16}. Fu \cite{fu} manually compared the performance of 7 translation services and found Google Translate to be one of the best English-Chinese translations. We therefore opted to use Google translation services for this series of experiments.

\subsubsection{Translation as a Preprocessing Step}
Our basic approach uses an NMT (Google Translation services) to transform all artifacts in our dataset into mono-lingual (English) representations. This is accomplished by translating the documents sentence by sentence. Artifacts were first split into sentences using the pre-trained sentence tokenizer provided by NLTK's PunktSentenceTokenizer \cite{DBLP:conf/acl/Bird06}. 
Regular expressions were then used to identify bilingual sentences. In our case, both English and Chinese used within an artifact are represented with UTF-8 encoding,  regular expressions capture non-English sentences by checking the encoding of their characters.  Finally, each of the bilingual sentences were processed by Google Translate to generate their English counterparts. These were then used to replace the bilingual sentences in their relevant artifacts.  As a result, the IBA dataset was transformed into an English mono-lingual dataset.

Although Markus \etc \cite{muhr2010external} suggested the use of token-by-token translation, we opted for sentence level translation for several reasons.  First, the sentence-level approach allows words to be considered within context and thereby to better retain their semantics following the translation. Furthermore, Google Translation Service is capable of handling intermingled terms and phrases within a sentence automatically, as it leverages a single NMT model to translate between multiple languages even in cases where a sentence contains intermingled phrases \cite{johnson2017google}. Taking the commit message in Table~\ref{tab:commit_example} as an example, Google's sentence level translation will generate a result such as "PagerUtils offset bug, when offset needs to be modified to 0, the value is incorrect", while token level translation will produce a sentence such as "PagerUtils offset of bug, when offset need modify for 0 time, value incorrect". In this case, the token level translation distorted the sentence semantics as it translated the Chinese phrases without fully understanding their context. 
Our approach is also more efficient and cost-effective than document-level translation, as it significantly reduces the volume of data submitted to translation service by removing sentences written in pure English. This is important as Google Translate charges more money and responds more slowly on a larger text corpus. 

\begin{figure*}[t!]

    \label{fig: exp1_bar}
    \begin{subfigure}{.32\linewidth}
        \centering
        \includegraphics[width=\linewidth]{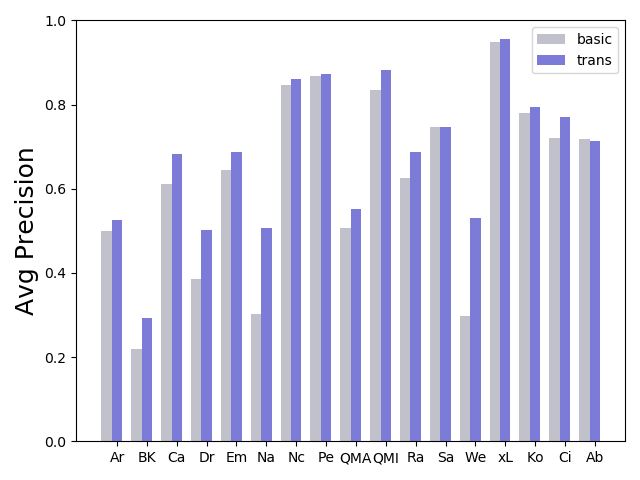}
        \caption{VSM. Average AP score 0.62 (VSM) and 0.68 (VSM$_{tr}$)}\label{fig:exp1_vsm}
    \end{subfigure}
        \hfill
    \begin{subfigure}{.32\linewidth}
        \centering
        \includegraphics[width=\linewidth]{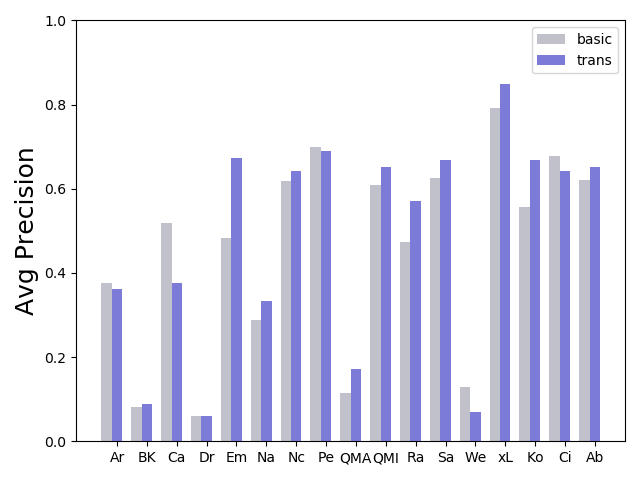}
        \caption{LDA. Average AP score 0.45 (LDA) and 0.48 (LDA$_{tr}$)}\label{fig:exp1_lda}
    \end{subfigure}
       \hfill
    \begin{subfigure}{.32\linewidth}
        \centering
        \includegraphics[width=\linewidth]{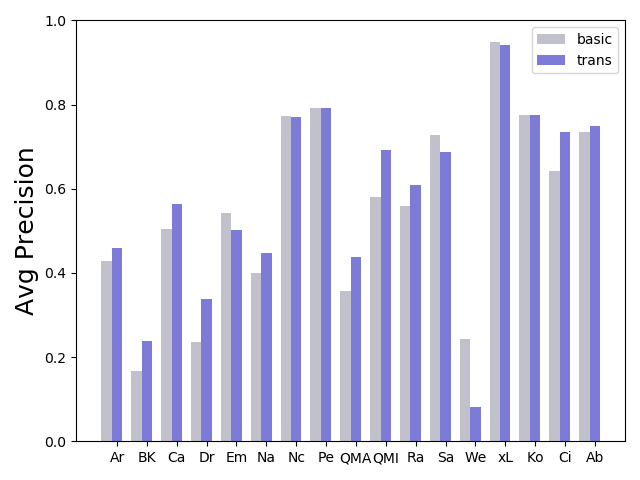}
        \caption{LSI. Average AP score 0.55 (LSI) and 0.58 (LSI$_{tr}$)}\label{fig:exp1_lsi}
    \end{subfigure}
    \caption{ AP scores for three basic trace models, with and without Google Translate, for 17 IBA datasets. Overall best results are observed for the Vector Space Model (VSM).}
    \label{fig:exp1_bar}
    \vspace{-12pt}
\end{figure*}

\begin{figure}[tbh!]
    \centering\includegraphics[width=\linewidth]{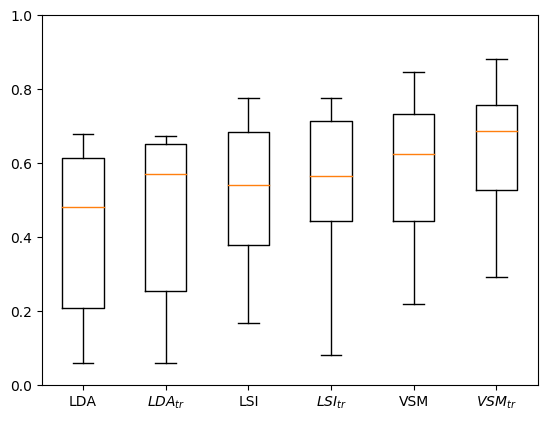}
            \vspace{-12pt}
    \caption{AP score distribution achieved when applying three models for 14 Chinese IBA datasets }
    \vspace{-12pt}
    \label{fig:exp1_box}
\end{figure}

\subsection{Evaluating NMT as a preprocessing step}
We utilized VSM, LSI, and LDA models, to automatically generate trace links for both sets of artifacts (i.e., the original IBA artifacts and the translated mono-lingual ones).  We then applied the time-constraints described in Section \ref{sec:datasets} to filter the resulting links.

\subsubsection{Metrics}
 Results were evaluated using  the commonly adopted average precision (AP) metric \cite{DBLP:conf/icse/ShinHC15}. AP evaluates the extent to which correct links are returned at the top of a list of ranked links. In this sense, it is considered more insightful than recall and precision metrics which simply measure whether a link is found within a set of candidate links. AP is calculated as follows:
\begin{equation}
    \label{eq:AP}
    AP = \frac{\sum_{i=1}^{n} Precision(i) \times rel(i)
    }{|true \  links|} 
\end{equation}
where $n$ is the number of candidate links, and rel(i) is an indicator function returning 1 if the link at $i_{th}$ ranking is a true link, otherwise return 0. Precision(i) is the precision score at the cut-off of $i_{th}$ ranking in the returned result. The $|true links|$ denominator refers to the total number of true links, meaning that we evaluate average precision for all true links and report AP scores at recall of 100\%.

\subsubsection{Results and Analysis}
To address RQ2 we compared the AP scores produced for each of the models, with and without Google Translate, for all 14 IBA datasets. The basic models are labeled VSM, LDA and LSI and the corresponding models using NMT are labeled VSM$_{tr}$, LDA$_{tr}$, and LSI$_{tr}$.
Detailed results for each project are reported in Fig. \ref{fig:exp1_bar}, and aggregated results across all projects are presented in Fig. \ref{fig:exp1_box}. We used the Wilcoxon signed-rank test \cite{woolson2007wilcoxon} to measure whether the use of translation statistically improved the performance for each technique. This is a standard test used to evaluate tracing algorithms due to the non-normal distribution of data points. We tested 14 pairs of AP scores achieved from the 14 datasets, with and without translation, using Scipy's \cite{scipy} Wilcoxon test function and adopted the standard significance threshold of 0.05. 

Results showed that VSM$_{tr}$ outperformed VSM with statistical significance (W=2, P = 0.001). On the other hand, in the cases of LSI vs LSI$_{tr}$ and LSI, (W = 34,P = 0.079) and for LDA$_{tr}$ and LDA (W = 43, P = 0.113) there was no statistically significant difference, given that in both cases, the p-values were above the significance threshold. These results indicate that translation improves performance in the case of VSM, but not necessarily for LSI and LDA, quite possibly because both of these techniques create topic models which can include terms from both languages.  As we can see in Fig. \ref{fig:exp1_box}, both LDA$_{tr}$, and LSI$_{tr}$ have higher medians, Q1, and Q3 values than their non-translation versions but a lower minimum value. It indicates that in certain cases, translation can degrade the performance of the tracing algorithm instead of improving it. This phenomenon is highlighted in Fig. \ref{fig:exp1_bar}, where we observe that in most projects, the `trans' version of LDA and LSI have a higher AP score, but there are a few exceptions in which the basic trace models perform better. This result also confirms previous findings that VSM often outperforms LDA and LSI in various mono-lingual tracing scenarios \cite{DBLP:conf/iwpc/OlivetoGPL10,DBLP:conf/sigsoft/LoharAZC13}, our experiment therefore extends this finding to the IBA domain. Given that VSM tends to outperform LSI and LDA on software engineering traceability datasets, it is particularly significant that VSM$_{tr}$ provides additional improvements. 
These results show that in the case of VSM, the translational preprocessing step improves accuracy of the results, and further imply that the presence of bilingual artifacts has a negative effect on traceability results. 

\subsubsection{Translation Related Pitfalls}
\label{sec:trans_limitaiton}
A careful analysis of individual tracing results unearthed three scenarios in which translation negatively impacted the results. \vspace{-8pt}\\


\noindent{\bf Scenario 1: }  A single Chinese term occurring in both source and target artifacts is sometimes translated into different English terms by NMT due to differing contexts. For example, before the translation step, a foreign term \begin{CJK*}{UTF8}{gbsn}‘启动’\end{CJK*} appeared as a verb written in Chinese and was shared by both source and target artifacts. It resulted in high similarity scores between the artifacts.  Following the translation step, the term was transformed to `start' in the source artifact and as `startup' in the target artifact. Neither VSM, LDA, or LSI captured the semantic similarity between these terms. As observed in our earlier study of usage patterns, English sentences seldom contain intermingled Chinese terms; therefore we primarily observed this scenario when tracing between pairs of artifacts written in Chinese (e.g., issues and code comments).\vspace{-8pt}\\

\noindent{\bf Scenario 2: } A relatively specific term in Chinese, translated into a common English term could introduce noise and subsequently increase the number of false positive links. As summarized in RQ1, commits with $CD_C$ tags may contain Chinese terms representing SQL query keywords or UI widget labels. Although these terms are usually sparse with respect to the size of the source code, they serve as strong indicators of true links when directly referenced in an issue discussion. \vspace{-8pt}\\

\noindent{\bf Scenario 3:} Unique phrases that describe specific features in Chinese, are eliminated or weakened by the translation step. We observed that artifact creators appeared to deliberately reference Chinese language content from other artifacts as signature features, indicating that two artifacts are closely related. Translation may inadvertently weaken the association by translating distinctive Chinese terms into common English words, some of which might even be removed as common stop words. This scenario was observed in some artifacts tagged as TAG$_C$ tag.

Despite these limitations, our results show that adding a translation step into the tracing workflow generally improves results with statistical significance as previously discussed. The main reason that NMT impairs tracing results is that similar terms can become distant from each other following translation. We therefore leveraged word embedding as a semantic layer to enable those terms to be mapped closer together in the multi-dimensional space in order to improve the tracing results.


\section{Generalized Vector Space Models}
\label{sec:approach}
Even though the results from our first experiment showed that integrating an NMT approach as a preprocessing step can improve trace accuracy in IBA datasets, we also identified three translation related pitfalls. To address these problems we propose a novel approach that combines the use of the Generalized Vector Space Model (GVSM) with both cross-lingual and mono-lingual word embeddings. We specifically address the following research questions which are described more extensively in subsequent sections:

\begin{itemize}[leftmargin=*]
\item {\bf RQ3:} To what extent does cross-lingual word embedding improve GVSM performance for tracing across IBA datasets?
\end{itemize}
\begin{itemize}[leftmargin=*]
\item {\bf RQ4:} Which of the tracing techniques presented in this paper perform best on IBA datasets?
\end{itemize}

\subsection{Brief Overview of GVSM}
GVSM \cite{DBLP:conf/sigir/WongZW85} is a generalization of the traditional VSM, designed specifically for information retrieval tasks. One of the known weaknesses of VSM is that terms only contribute to the similarity score if they appear in both source and target artifacts.
GVSM directly addresses this problem by creating an additional layer of latent space, in which the distance between the terms is determined by their semantic similarity. Given a collection of artifacts $a_i \in A_S\cup A_T$, $a_i$ is vectorized using the standard VSM approach such that $a_i=\{w_1,w_2,\dots,w_n\}$ where $w_n$ are the weights for the terms in artifact $a_i$. Considering the vocabulary $V=\{t_1,t_2,\dots,t_N\}$ composed by all the terms in artifacts, the pairwise term-correlation can be represented as a correlation matrix G of $N \times N$ shape, where $sim(t_i,t_j)$ is the semantic relevance for term $t_i$ and $t_j$ 
\begin{equation}
    G = \begin{vmatrix}
        sim(t_1,t_1) & sim(t_1,t_2) & \dots\\
        \vdots & \ddots & \\
        sim(t_N,t_1) & & sim(t_N,t_N)\\
    \end{vmatrix}
\end{equation}
In GVSM the similarity between two artifacts is then calculated as follow:
\begin{equation}\label{GVSM_EQ}
    Similarity(a_i,a_j) = \frac{a_i^T \cdot G \cdot a_j}{\sqrt{a_i^T\cdot G \cdot a_i}\sqrt{a_j^T\cdot G \cdot a_j}}
\end{equation}

GVSM has been effectively used for to support bilingual tasks related to document clustering \cite{DBLP:conf/ijcnlp/TangXZLZ11}, query processing\cite{DBLP:journals/is/WongZRW89}, and text retrieval\cite{DBLP:conf/eacl/TsatsaronisP09}.

\subsection{Word Embedding (WE)}
\label{sec:word_embedding}
For text processing purposes, all terms need to be represented in a format that is conducive to machine learning. Many traceability algorithms  encode terms as distinct symbols without considering their semantic associations. A common approach is to use `one hot encoding' in which terms are represented as vectors in a high dimensional orthogonal space in which each term is a dimension of the vector space. In contrast, word embeddings transform the word representation from a high dimensional orthogonal space to a low dimensional non-orthogonal space. The distribution of vectors within the space varies depending on the specific approach taken. 
The use of word embedding has achieved significant success for addressing NLP challenges in  domains such as ad-hoc information retrieval\cite{DBLP:conf/ecir/AlmasriBC16}\cite{DBLP:conf/sigir/VulicM15}, bug localization\cite{DBLP:conf/bigdataconf/YeQM15},  question
answering\cite{DBLP:conf/acl/DhingraZFMC16} and also trace link recovery\cite{DBLP:conf/apsec/ZhaoCS17}\cite{DBLP:conf/icse/0004CC17}.

\subsubsection{Mono-lingual Neural Word Embedding}
\label{sec:mono_wv}
Word embeddings typically represent terms from a single language, and have the intuitive assumption that terms with similar distribution patterns have closer semantic meaning than those with dissimilar distributions \cite{harris54}. This means that term vectors tend to form clusters according to their semantic affinity. Mono-lingual neural word embeddings (MNWE) leverage the context of each term and include the Skip-Gram model and Continuous Bag Of Words (CBOW) model \cite{DBLP:journals/corr/abs-1301-3781}. While both of these models are built upon simple three-layer neural networks, the Skip-Gram model makes predictions of surrounding terms based on the current terms while the CBOW model predicts a single term using all surrounding terms as context. In our study, we adopt pre-trained mono-lingual word embedding that are trained on Common Crawl dataset with enhanced CBOW model\cite{mikolov2018advances}. Vectors in such a space have 300 dimensions.

\subsubsection{Cross-lingual Word Embedding (CLWE)}
\label{sec:cl_wv}
Cross-lingual word embeddings project the vocabulary from two or more distinct languages into a single vector space. As with mono-lingual embeddings, a reasonable cross-lingual embedding model should be capable of organizing term vectors in an embedding space according to their semantic affinities. Cross-lingual embeddings can therefore serve as a semantic bridge between different languages, and can be used to support diverse cross-lingual NLP tasks such as machine translation, cross-lingual IR, and cross-lingual entity linking. 
Various techniques have been explored for aligning multiple mono-lingual vector spaces \cite{Ruder2017ASO}. Techniques based on word-level alignment tend to leverage cross-lingual lexicons to supervise the vector mapping.  With this approach, mono-lingual word embeddings are first created for both languages, and then the unseen term vectors for both languages are transformed using a trained mapping model. Researchers have explored other approaches for relaxing word-level alignment constraints, for example by leveraging alignment information at the sentence level \cite{DBLP:conf/icml/GouwsBC15} or even the document level \cite{DBLP:conf/eacl/Vulic17}, or entirely foregoing supervision \cite{DBLP:journals/corr/abs-1809-02306}, in order to extend the use of cross-lingual word embedding to additional scenarios.\\ \vspace{-8pt}

For our experiments we utilized  \emph{relaxed cross-domain similarity local scaling} (RCSLS), which is based on word-level alignment \cite{DBLP:conf/emnlp/JoulinBMJG18}. We selected RCSLS for two reasons.  First it has been shown to deliver the best overall performance in comparison to other state-of-the-art techniques across 28 different languages \cite{DBLP:conf/emnlp/JoulinBMJG18}, and second, pre-trained models with support for 44 languages is available from Facebook \cite{joulin2018loss}.  Facebook trained their model  using  Wikipedia documents, and leveraged the MUSE library\cite{conneau2017word} which includes bilingual dictionaries for over 110 languages and mono-lingual word embeddings for 157 languages including English and Chinese. Vectors in cross-lingual embedding space also have a dimension of 300.

\subsection{Combining GVSM with Cross-Lingual WE}
\label{sec:combine_gvsm_we}
Prior work has already investigated the application of GVSM for cross-lingual information retrieval tasks in other domains. For example, Tang \etc \cite{DBLP:conf/ijcnlp/TangXZLZ11} proposed CLGVSM which exploited semantic information from (1) a knowledge base e.g. HowNet (Xia et al., 2011), (2) statistical similarity measures, e.g cosine similarity of term vector covariance (COV), and (3) a bilingual dictionary which contains the translation probability between terms.
Another branch of study attempts to leverage Cross-Lingual Word Embedding to address cross-lingual information retrieval tasks.  Vulic et al. \cite{DBLP:conf/sigir/VulicM15} proposed a model known as cross-lingual information retrieval (CLIR) which directly leverages the distributed representation of cross-lingual vocabulary to accomplish document embedding (DE). Given documents represented by term vectors $d = \{\Vec{t_1},\Vec{t_2},\dots,\Vec{t_n}\}$ where $\Vec{t_i}$ is the vector representation of terms, a document vector $\Vec{d}$ can be created by merging the term vectors with simple addition. The self-information of the terms \cite{Cover:2006:EIT:1146355}, e.g. frequency of terms within a document, can be combined to weight the vectors. The final representation of a document is given as:
\begin{equation}
    \Vec{d} = w_1 \Vec{t_1} + w_2 \Vec{t_2} + \dots + w_n\Vec{t_n}
\end{equation}
This method, referred to as ADD-SI, projects the documents into the same vector space of terms so that the document affinities can be evaluated using distance measures such as cosine similarity. 
However, we could not find any publications describing the combined use of both GVSM and Cross-Lingual Word Embedding. We replaced the cross-lingual knowledge base in CLGVSM with (Cross-Lingual) Word Embedding, because knowledge tended to be domain-specific and costly to construct for individual software project domains.
We therefore propose three different techniques for combining GVSM with Word Embeddings.  

\begin{table}[t!]
    \begin{footnotesize}
    \setlength{\tabcolsep}{0.65em}
	\centering
	\caption{ The acronyms and details of the three GVSM and word embedding integrated methods: WE=Mono-lingual word embedding, CL=Cross-lingual word embedding, TR=Google translation to English}
	 \small\addtolength{\tabcolsep}{-4pt}
        \begin{tabular}{|l|c|c|c|c|m{5.4cm}|}\hline
        {\bf Abbr} & GVSM & WE & CL & TR& {\bf Description}\\ \hline
         CLG& $\blacksquare$ & &$\blacksquare$&& Uses Cross-Lingual word embedding with GVSM. Inputs a bilingual dataset to the model\\ \hline
        WEG& $\blacksquare$& $\blacksquare$&&&Uses reduced size English Word embedding with GVSM. Inputs a bilingual dataset to the model. $WEG^{*}$ variant uses full-sized English word embedding.\\ \hline
        $WEG_{tr}$ &$\blacksquare$& $\blacksquare$&&$\blacksquare$& Uses reduced size English word embedding with GVSM. Uses Google Translate to preprocess IBA data. Inputs  resulting mono-lingual dataset to model.  $WEG^{*}_{tr}$ variant uses full-sized English word embedding.\\ \hline

        \end{tabular}
	\vspace{-6pt}
	\label{tab:gvsm_models}
	\end{footnotesize}
\end{table}

\subsubsection{Cross-Lingual Word Embedding with GVSM (CLG)}
Our first approach uses a modified cross-lingual word embedding based on GVSM. As shown in Equation \ref{GVSM_EQ}, a GVSM model is composed of TF-IDF vectors $a_i$ and a semantic relatedness matrix $G$. The semantic relatedness matrix $G$ can be constructed using external knowledge\cite{DBLP:conf/eacl/TsatsaronisP09, DBLP:conf/ijcnlp/TangXZLZ11} (e.g HowNet, WordNet) to evaluate term relatedness based on their distance in the knowledge network; or using statistical models that predict the probability of term co-occurrence\cite{DBLP:conf/sigir/WongZW85}.  In the first approach, the size of the semantic relatedness matrix is constrained by the vocabulary of the knowledge base. This is a critical limitation for trace link recovery, as software artifacts tend to include a large number of technical terms which are not commonly found in general purpose knowledge sources.  Statistical approaches therefore fall far short of capturing the true semantics of these terms. However, these weaknesses can be addressed using word embeddings.

Given an IBA dataset with primary and foreign language vocabulary $L_p = \{t_{p_1}, t_{p_2}, \dots, t_{p_m}\}$, $L_f = \{t_{f_1}, t_{f_2}, \dots, t_{f_n}\}$, the mono-lingual vector for $t_{p_i}$ and $t_{f_i}$ is represented as $x_i,z_i \in \mathbb{R}^d$ where d refers to the dimension of the vector space. As previously discussed, the RCSLC model is capable of projecting vectors from two separate spaces into a shared latent space by learning an internal translation matrix W with the supervision of bilingual lexicons. With this translation matrix W, the vectors can be projected as $Wx_i \ and \ Wz_i$. The vocabulary vector space of a given IBA dataset is then represented as:
\begin{equation}
\label{eq:vocab_vec}
V_S= \{Wx_1,\dots, Wx_m,Wz_1,\dots,Wz_n\}
\end{equation}
As the vectors in RCSLC are $l-$2 normalized\cite{DBLP:conf/emnlp/JoulinBMJG18}, the semantic relevance matrix $G$ can be created through the simple dot product of $V_S$ and $V_S^T$.
The GVSM formula shown as equation \ref{GVSM_EQ} can be transformed into the following:
\begin{equation}
\label{eq:gvsm_sim}
    Similarity(a_i,a_j) = \frac{a_i^T \cdot V_S \cdot V_S^T \cdot a_j}{\sqrt{a_i^T \cdot a_i}\sqrt{a_j^T\cdot a_j}}
\end{equation}
In our case, $V_S$ is the pre-built vector space provided by FastText library as described in Sec. \ref{sec:word_embedding}

\begin{figure*}[t!]
    \centering\includegraphics[width=\linewidth]{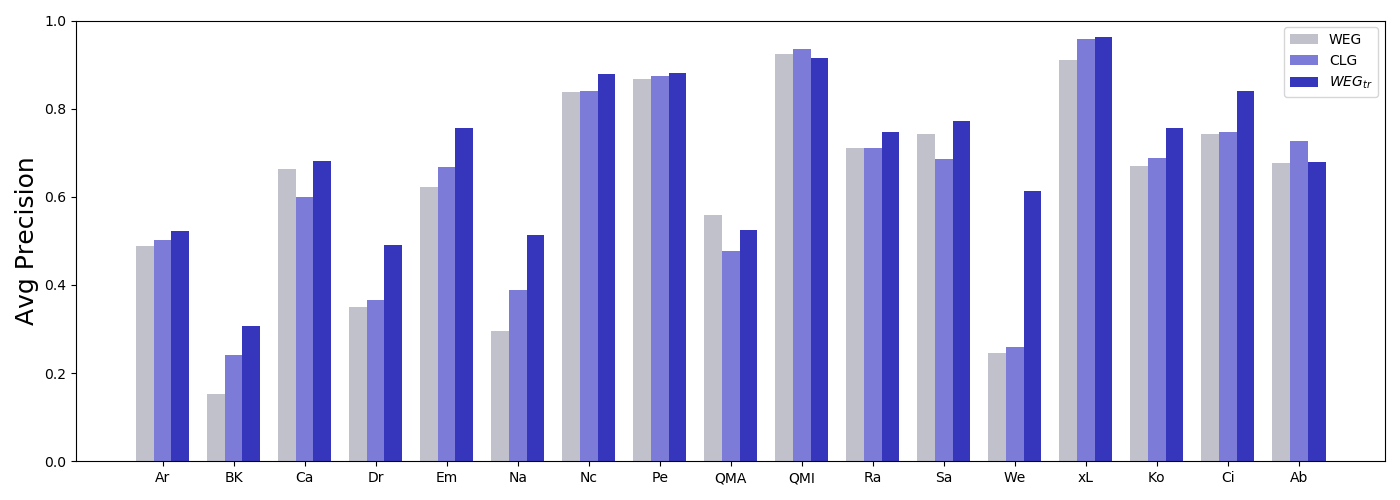}
       \vspace{-16pt}
    \caption{Average performance of three primary GVSM models for all 17 IBA datasets}
    \label{Fig:exp2_gvsm_performance}
\end{figure*}

\begin{figure}[tbh!]
    \centering\includegraphics[width=\linewidth]{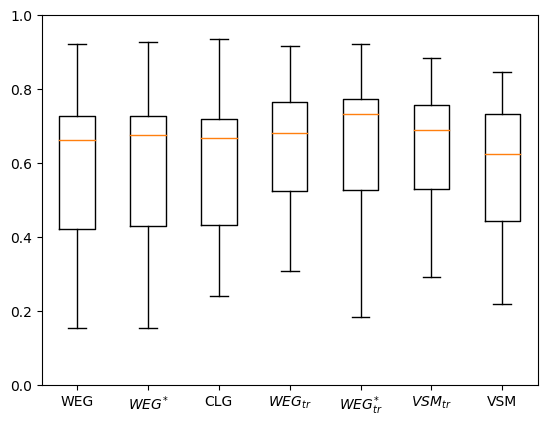}
    \caption{A comparison of the best basic model (VSM$_{tr}$) against all three GVSM-based models. WEG and WEG$_{tr}$ with full size English embedding are represented as WEG$^{*}$ and WEG$^{*}_{tr}$} 
    \vspace{-12pt}
    \label{Fig:exp2_gvsm_performance_detail}
\end{figure}

\subsubsection{Transforming CLG to a Mono-lingual Tracing Task}
For experimental purposes we also explored a mono-lingual version of CLG. The intrinsic difference between mono-lingual and bilingual trace tasks lies in the dataset vocabulary. As shown in Equation \ref{eq:vocab_vec}, the vocabulary vector space $V_s$ of IBA dataset is composed from two types of vectors 1) term vectors projected from the foreign language space and 2) term vectors projected from English space. For mono-lingual tracing tasks, the vocabulary vector space $V_{S^{'}}$ contains term vectors of only one language; however, by simply substituting the $V_s$ with $V_{S^{'}}$ in Equation \ref{eq:gvsm_sim} we can migrate CLG to address mono-lingual tracing tasks. This can be accomplished by training and applying a word embedding model with mono-lingual text corpus. We name this mono-lingual model the `Word Embedding GVSM' (WEG) to distinguish it from CLG.
\subsubsection{NMT preprocessing with Mono-lingual Trace Task}
As we described above, WEG is the mono-lingual version of CLG in which cross-lingual word embedding is replaced with an English mono-lingual embedding. We also propose a third approach which combines WEG with NMT to extend its ability to trace IBAs. We followed the same approach used in our initial experiments with  VSM$_{tr}$, LDA$_{tr}$ and LSI$_{tr}$, by using Google Translation services to translate the IBA datasets back into English mono-lingual datasets before running WEG. We refer to this method as \emph{WEG$_{tr}$} to distinguish it from the other two GVSM models.

\subsection{Experiment}
\label{sec:rq2}
All three GVSM models (i.e., CLG, WEG and WEG$_{tr}$) shown in Table \ref{tab:gvsm_models} were applied against our experimental datasets. 
However, due to different amounts of training data available, the size of the cross-lingual embedding tends to be smaller than the mono-lingual word embedding. To make a fair comparison between the techniques of using mono-lingual and cross-lingual embeddings, we randomly sampled the vocabulary of the mono-lingual word embedding to reduce its size. The full mono-lingual embedding included 2,519,371 records, while the cross-lingual embedding and the down-sized mono-lingual embedding included only 332,648 records.  However, in reality, it is far easier to construct a large mono-lingual word embedding, and therefore we wanted to see how WEG$_{tr}$ and WEG performed when allowed to use the fully available embedding data.  We therefore also include these results, labeled as WEG$^{*}_{tr}$ and WEG$^{*}$ respectively. Finally, as an additional point of comparison, we include both VSM and VSM$_{tr}$ from our earlier experiment. Results are reported in Figures \ref{Fig:exp2_gvsm_performance} and \ref{Fig:exp2_gvsm_performance_detail}.

\subsubsection{RQ3: Analysis of Cross-lingual Word Embedding}
To address our research question ``To what extent does cross-lingual word embedding improve GVSM performance for tracing across IBA datasets?'' we explore the difference between CLG, WEG, and WEG$_{tr}$ models.

By comparing average precision achieved for the 14 Chinese datasets as reported in Figure \ref{Fig:exp2_gvsm_performance} we observe that in 12 out of 14 cases, WEG$_{tr}$ was the winner. Of the remaining two cases, CLG and WEG each won once.  Furthermore, as reported in  Figure \ref{Fig:exp2_gvsm_performance_detail}, WEG$_{tr}$ has a significantly higher median, Q1, and Q3 value than either mono-lingual WEG or CLG. Applying full size word embedding to WEG$_{tr}$ further improves the performance by increasing the median value of the results distribution. This indicates that combining WEG with NMT can effectively improve the tracing performance. When comparing WEG and WEG$^{*}$, we observed that increasing the embedding size for mono-lingual WEG has little impact on model performance; however, this contrasts with the marked improvement observed when using an increased English embedding size on WEG$_{tr}$, reinforcing our conjecture that the vocabulary mismatch introduced by IBA has a clear negative impact upon trace performance. 

To determine if it would be possible to avoid the costs of building or contracting a translation service such as Google Translation services, we also compared the mono-lingual and cross-lingual approaches (i.e., WEG vs. CLG) without the benefit of translation. In this case, we observe that CLG outperforms WEG in 10 out of 14 Chinese projects, achieves equivalent performance in one project, and underperforms in 3 projects. However, an analysis of results in Figure \ref{Fig:exp2_gvsm_performance_detail} shows that its median, Q1, and Q3 values in comparison to other models, show that it does not statistically outperform WEG.

We therefore answer RQ3 by stating that the cross-lingual word embedding failed to outperform either of the mono-lingual word embedding approaches based on the available resources, and that the use of a preprocessing translation step followed by the use of GVSM with mono-lingual word embedding was clearly superior.

\subsubsection{RQ4: Comparison of all models}
Finally, to address our research question ``Which of the tracing techniques presented in this paper perform best on IBA datasets?'' we compare VSM$_{tr}$, LDA$_{tr}$, and LSI$_{tr}$ with our new GVSM-based techniques. As Fig. \ref{fig:exp1_box} and Fig. \ref{Fig:exp2_gvsm_performance_detail} report VSM$_{tr}$ and WEG$^{*}_{tr}$ are observably the best models. We compared AP scores achieved for these two models for all 14 Chinese datasets, against each of the other models using the Wilcoxon signed-rank test and Cohen's d effect size. P-value of Wilcoxon signed-rank test are reported in Table \ref{tab:wilcoxon}, show that both VSM$_{tr}$ and WEG$^{*}_{tr}$ are statistically significant better than other models given the P-values are all below 0.05 with effect size ranging from 0.3 to 0.9 indicating a "medium" to "large" effect. However, a similar comparison of WEG$^{*}_{tr}$ and VSM$_{tr}$ returns a P-value 0.0615 and effect size of 0.09, meaning that neither technique is significantly better than the other even though Fig. \ref{Fig:exp2_gvsm_performance_detail} shows that WEG$^{*}_{tr}$ has a higher maximal, Q1, Q3 than VSM$_{tr}$.  
 \begin{table}[t!]
    \centering
     \caption{P-value of wilcoxon signed-rank test.}
    \begin{tabular}{ccccc}
    \toprule
                   &$WEG^{*}$& CLG &$LDA_{tr}$&$LSI_{tr}$\\\hline
    $WEG^{*}_{tr}$ &    .001    &  .003  &  0  .000      &     .001 \\
    $VSM_{tr}$     &    .019    &  .010  &  0   .000    &     .001 \\ \hline
    \end{tabular}
   
    \label{tab:wilcoxon}
\end{table}

\section{Extension to Other Languages}
\label{sec:analysis}
While our focus was on Chinese-English language projects, we also included one project from each of three additional languages in our experiments as a preliminary proof of concept.  These projects were Korean, Japanese, and German -- all combined with English. In all cases, including the German language, we were able to use Unicode to identify its presence in English sentences. While we were able to identify language occurrences in our study (i.e., Chinese, Japanese, Korean, and German) from English using Unicode, we will need to adopt more diverse approaches (e.g., Python's langdetect project), for language detection\cite{faq}. Traceability results for these projects are reported throughout the paper (shown on the right hand side of Table \ref{tab:data_status}, and the graphs of Figures Fig.~\ref{Fig:exp2_gvsm_performance} and Fig.~\ref{fig:exp1_box}). They show that for both Asian languages (Korean and Japanese) our conclusion derived from the Chinese datasets is still valid, while for the European language (German), the CLG model outperformed WEG$_{tr}$.  We leave a deeper analysis of this observation to future work.

\section{Threats to Validity}
\label{sec:threats}
There are several threats to validity in this study. First, we used Google as our black-box translator.  As the vocabulary selection strategy of the NMT has a direct impact on the final trace link quality, results could be different if other types of machine translation methods are applied. However, we chose Google translation as it has been empirically shown to deliver high quality  translations across numerous languages.
Another important threat is that the training material used for CLG was composed of general documents from Wikipedia and did not include domain specific corpora. We made this decision in order to deliver a more generalized solution, and because collecting a domain specific corpus for all 17 projects would have been prohibitively expensive. CLG might perform better than WEG if a domain-specific corpus of technical documents had been available and used for training purposes.  An external threat was introduced by limiting the raw artifacts to the coverage area of intermingled links to alleviate the link sparsity issue (see Table \ref{tab:data_status_2}). This enabled us to focus on the IBA-impacted traces, but reduced the number of participating artifacts, thereby potentially inflating AP scores for all results.  Also, we did not yet explore the impact of embedding size on CLG. On the other hand, our experiments showed that  WEG$_{tr}$ still outperformed CLG, when equal sized embeddings were used. The reality is that larger mono-lingual embeddings are more readily available, and should be pragmatically leveraged. We leave experimentation with different sized embeddings to future work. Finally, we extensively evaluated the results on Chinese-English projects, but study is required to generalize our finding to other languages.
\section{Related Work}
\label{related_work}
Prior work related to cross-lingual translation has already been extensively described throughout the paper. In this section we therefore focus on the very limited body of work that exists in the use of multiple languages in the software development environment. This phenomenon tends to occur in two primary settings -- first in organizations in which English is not the primary language of many stakeholders, but is the language of choice for supporting the development process; and second, in global software development environments with geographically dispersed teams speaking multiple languages. As Abufardeh \etc \cite{5488634} points out, this kind of geographical localization for development teams is a critical element of the success of multi-national projects.

Multi-lingual problems in global software development (GSD) have been identified and discussed by previous researches. Although English is widely accepted as an official languages in most international IT corporations, the practice of utilizing a second language is quite common. For example, Lutz \cite{DBLP:conf/icgse/Lutz09} investigated the issue in Siemens'  English-German work space and pointed out that utilizing English as the primary language for non-native speakers can lead to misunderstandings in both oral and written communication. 

Researchers have proposed different methods to address the multi-lingual linguistic problem. One branch of studies has focused on developing  best-practices\cite{DBLP:journals/cacm/KrishnaSW04} to enhance the work quality and efficiency, while others have proposed using  machine translation as a solution for minimizing the misunderstanding brought by multi-lingual environment \cite{DBLP:conf/icgse/CalefatoLP11, DBLP:conf/cscwd/MoulinSFWM09, DBLP:conf/icse/Cleland-HuangCDGHKLMPSZABEM11}. Our approach best fits into this second category as we have observed the problem of multi-lingual language use in software artifacts and have applied diverse machine-learning solutions to compensate for in the traceability task.

\section{Conclusion }
\label{conclusion}
The work in this paper was motivated by the needs of our industrial collaborators who were seeking enterprise-wide traceability solutions across software repositories containing artifacts written in a combination of English and Chinese. 

This paper has made several contributions.  First, it explored the use of intermingled Chinese and English terms across 14 different projects and identified common usage patterns. It then showed that using a preprocessing translation step in IBA projects in conjunction with three commonly used trace models improved accuracy of the generated trace links.  This clearly indicates that the multi-lingual problem must be addressed for traceability in IBA datasets. We then proposed three GVSM based methods which leveraged the strength of word embedding to address the IBA vocabulary issue. Our experiment results showed that, WEG$^{*}_{tr}$ can outperform NMT combined classic trace models and two other GVSM based methods proposed by us. In the cases where NMT may not be available (due to costs of an external service provider), we propose CLG and WEG as viable alternatives, because it is easier, more effective, and less costly to train a word embedding based model than an NMT translator. Furthermore, an internally trained CLG and WEG model could potentially include domain-specific terminology, thereby potentially boosting its performance.

\section*{Acknowledgements}
The work described in this paper is funded by NSF grant CCF-1649448

\balance
\bibliographystyle{ACM-Reference-Format}      
\bibliography{reference,trace}

\end{document}